\documentclass[10pt,journal,twocolumn]{IEEEtran}
\usepackage{cite}
\usepackage{amsfonts}
\usepackage{graphicx}
\usepackage{amsmath}
\usepackage{mathtools} 
\usepackage{times}
\usepackage[bookmarks]{hyperref} 

\DeclareMathOperator*{\E}{E}
\newcommand{\mytilde}{\raise.17ex\hbox{$\scriptstyle\mathtt{\sim}$}}

\begin{document}
\title{On the Error Performance of \\ Systematic Polar Codes}

\author{
Liping~Li,~\IEEEmembership{~Member,~IEEE,} Wenyi~Zhang,~\IEEEmembership{~Member,~IEEE}, Yanjun Hu,~\IEEEmembership{~Senior Member,~IEEE}
\thanks{
Liping Li and Yanjun Hu is with the Key Laboratory of Intelligent Computing and Signal Processing of the Ministry of Education of China, and School of Electronics and Information Engineering, Anhui University, China (email: Liping Li's email: \href{mailto:liping_li@ahu.edu.cn}{\nolinkurl{liping_li@ahu.edu.cn}}, Yanjun Hu's email: \href{mailto:yanjunhu@ahu.edu.cn}{\nolinkurl{yanjunhu@ahu.edu.cn}}).}
\thanks{Wenyi Zhang is with School of Information Science and Technology, University of Science and Technology of China (email: \href{mailto:wenyizha@ustc.edu.cn}{\nolinkurl{wenyizha@ustc.edu.cn}}).}
}


\maketitle
\begin{abstract}
Systematic polar codes are shown to outperform non-systematic polar codes in terms of the bit-error-rate (BER)
performance. However theoretically the mechanism behind the better performance of systematic polar
codes is not yet clear. In this paper, we set the theoretical framework to analyze the performance of systematic
polar codes. The exact evaluation of the BER of systematic polar codes conditioned on the BER of
non-systematic polar codes involves in $2^{NR}$ terms where $N$ is the code block length and
$R$ is the code rate, resulting in a prohibitive number of computations for large block lengths.
By analyzing the polar code construction and the successive-cancellation (SC) decoding process,
we use a statistical model to quantify the advantage of systematic polar codes over non-systematic polar codes,
so called the systematic gain in this paper.
A composite model is proposed to approximate the dominant error cases in the SC decoding process.
This composite model divides the errors into independent regions and coupled regions, controlled
by a coupling coefficient. Based on this model, the systematic gain can be conveniently calculated.
Numerical simulations are provided in the paper showing very close approximations of the proposed model
in quantifying the systematic gain.

\end{abstract}

\section{Introduction}\label{sec_ref}

Polar codes are systematically introduced by Arikan in \cite{arikan_iti09}. It's shown
there that polar codes can achieve the capacity for symmetric binary-input discrete
memoryless channels (B-DMC) with a low complexity. The encoding and decoding process (with successive cancellation, SC)
can be implemented with a complexity of $\mathcal{O}(N \log N)$. The polarization
of $N$ channels is realized through two stages: channel combining and splitting.
Channels are polarized after these two stages in the sense that bits transmitting in these
channels either experience almost noiseless channels or almost completely noisy channels for a large $N$.
The idea of polar codes is to transmit information bits on those noiseless channels while fix
the information bits on those completely noisy channels. The fixed bits are made known
to both the transmitter and receiver.
The binary input alphabet in Arikan seminal work \cite{arikan_iti09} is
later on extended to non-binary input alphabet \cite{sasoglu_09,mori_itw10,pradhan_allerton11}.
The construction of polar codes have then been investigated and different procedures are
proposed \cite{mori_isit09, telatar_isit11, trifonov_itc12, vardy_polar} assuming the original $2 \times 2$ kernel matrix.
Polar codes based on the kernel matrices of size $l \times l$ are studied in \cite{korada_iti10}.
Polar codes have also been extended to different scenarios since then
\cite{telatar_itw10,vardy_iti11,telatar_iti12}.

The rate of polarization of polar codes is studied in \cite{arikan_iti09,arikan_isit09} without including
the effect of the code rate. In works \cite{urbanke_isit10,mori_isit10,hassani_iti13}, the
authors analyzed the polarization rate considering the effect of both the block length and the code rate.
The asymptotic behavior of polar codes reported in these works does not guarantee a good performance in practice when
a finite block length is applied.
In fact, the performance of polar codes with the SC decoding and finite block lengths are
not satisfactory \cite{urbanke_isit09}\cite{eslami_allerton10}.
Different decoding techniques are deployed to improve the
performance of polar codes \cite{urbanke_isit09, eslami_allerton10,arikan_icl08,guo_isit14,barry_icc13, vardy_isit11,niu_itc13,eslami_isit11}.
The authors of \cite{urbanke_isit09, eslami_allerton10,arikan_icl08, guo_isit14,barry_icc13} use
belief propagation (BP) in the decoding process in place of the SC decoding.
The list decoding procedure of \cite{vardy_isit11} and \cite{niu_itc13} involves
multiple paths instead of a single path as in the SC decoding process.
The concatenation of polar codes with LDPC codes are proposed in
\cite{guo_isit14} and \cite{eslami_isit11}  to further improve the performance of polar codes.
These techniques focus on the improvement in the decoding algorithms while keeping the original
coding process as in \cite{arikan_iti09}. The price paid in these improvements is the extra
decoding complexity.

Another direction to improve the performance of polar codes is
also introduced by Arikan in \cite{arikan_icl11} by using systematic polar codes. If we
denote $\underline{u}$ as a vector containing source bits and $\underline{x}$
as the corresponding codeword obtained by using the normal polar codes construction.
Note that in this paper we use non-systematic polar codes and normal polar codes
interchangeably without further notice. The basic idea
of systematic polar codes is to use some part of the codeword $\underline{x}$ to transmit information
bits instead of directly using $\underline{u}$ to transmit them. The
advantage of systematic polar codes is the low decoding complexity:
Systematic polar codes require only part of
the encoding process (involving only $0$s and $1$s) after the normal SC decoding is done.
This low complexity can be seen from the way $\underline{x}$ is estimated:
$\underline{\hat{x}} = \underline{\hat{u}}G$
where $\underline{\hat{u}}$ is the estimation of $\underline{u}$ from the normal SC decoding,
and $G$ is the generator matrix. In the rest of the paper, we call this indirect, two-step
(SC decoding then encoding) decoding
process of systematic polar codes the SC-EN decoding.

In \cite{arikan_icl11}, it's shown that systematic polar codes achieve better bit-error-rate
(BER) performance than normal polar codes. However, Arikan also noted in \cite{arikan_icl11}
that it's not clear why systematic polar codes achieve better BER performance than non-systematic
polar codes even with an indirect decoding procedure (the SC-EN decoding): first decoding $\underline{\hat{u}}$ then re-encoding
$\underline{\hat{x}}$ as $\underline{\hat{u}}G$. One would expect that any error in $\underline{\hat{u}}$
would be amplified from this re-encoding process $\underline{\hat{x}} = \underline{\hat{u}}G$.
However simulation results in \cite{arikan_icl11} as well as simulation results in this paper show
that with this two-step decoding procedure, systematic polar codes still achieve better BER performance
than non-systematic polar codes.

This paper studies the error performance of systematic polar codes with special focus on characterizing the
advantage of systematic polar codes over non-systematic polar codes.
We start by simplifying the general encoding process  of the systematic polar codes. This is done
through proving a theorem on the structure of the generator matrix.
Then we discuss the theoretical BER performance of systematic polar codes
conditioned on the BER performance of non-systematic polar codes.
The general form of this error prediction
involves in $2^{NR}$ terms which is prohibitive to compute for large block lengths $N$. It's then proven
that for two special cases we can theoretically predict the error rate of systematic polar codes.
To understand the general better behavior of systematic polar codes, we further
study the basic error patterns of non-systematic polar codes with the SC decoding.
A systematic gain is defined to describe the advantage of systematic polar codes over non-systematic polar codes.
A  composite model is proposed to approximate the mean effect (or the dominant effect)
of the error events. This composite model uses the fact that the errors in the SC
decoding process are coupled. A coupling coefficient is used to control the level of coupling between the errors.
This model facilitates the calculation of the systematic gain and can be used to predict the
performance of systems utilizing systematic polar codes.


Following the notations in \cite{arikan_iti09}, in the paper,
we use $v_1^N$ to represent a row vector with elements $(v_1,v_2,...,v_N)$. We also use
$\underline{v}$ to represent the same vector for notational convenience. Given a vector $v_1^N$, the
vector $v_i^j$ is a subvector $(v_i, ..., v_j)$ with $1 \le i,j \le N$. If there is a set $\mathcal{A} \in \{1,2,...,N\}$,
then $v_{\mathcal{A}}$ denotes a subvector with elements in $\{v_i, i \in \mathcal{A}\}$.

The rest of the paper is organized as follows. In Section
\ref{sec_background}, the background of systematic polar codes is introduced and a theorem on the
structure of systematic polar codes is proven.
The first part of Sec. \ref{sec_performance} provides a general theoretical formation
of the BER performance of systematic polar codes given the BER performance
of non-systematic polar codes. Two special cases are analyzed in this part whose BER performance can be
characterized. Section \ref{sec_error_pattern} studies the basic
error patterns and the first error distribution of non-systematic polar codes,
followed by the introduction of the systematic gain.
In Section \ref{sec_model}, we propose a coupling model which is used
to predict the BER performance of systematic polar codes.
Simulation results are given in Section
\ref{sec_numerical}. Concluding remarks are presented in Section \ref{sec_con}.


\section{Systematic Polar Codes}\label{sec_background}
For completeness, in the first part of this section, we restate the relevant materials on
the construction of normal polar codes and
systematic polar codes from \cite{arikan_iti09} \cite{arikan_icl11}. In the second part of this section
a theorem on the structure of the normal polar codes is provided which is used to simplify the
encoding of the systematic polar codes.

\subsection{Preliminaries of Non-Systematic Polar Codes}
\begin{figure*}
{\par\centering
\resizebox*{3.0in}{!}{\includegraphics{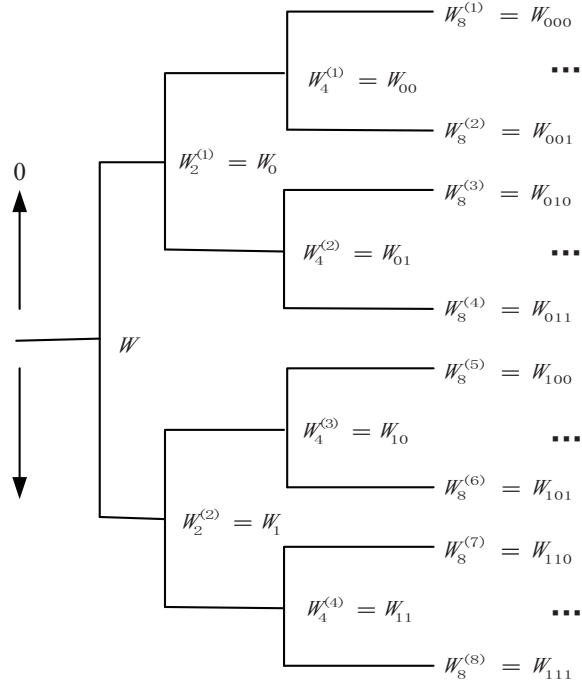}} \par}
\caption{The recursive channel transformation of polar codes. }
\label{fig_tree}
\end{figure*}

Let $W$ be any binary discrete memoryless channel (B-DMC) with a transition probability $W(y|x)$.
The input alphabet $\mathcal{X}$ takes values in $\{0,1\}$ and
the output alphabet is $\mathcal{Y}$.
Channel polarization is carried
in two phases: channel combining and splitting. Eventually, $N=2^n (n \ge 1)$
independent copies of $W$ are first combined and then split into $N$ bit channels
$\{{W_N^{(i)}}\}_{i=1}^N$. This polarization process has a recursive tree structure in \cite{arikan_iti09},
which we plot here for the ease of reference.
The $0$s and $1$s in Fig.~\ref{fig_tree} refer to the bit channels
$W'$ and $W''$ respectively in
the basic one-step channel transformation defined as $(W,W) \mapsto (W^{'},W^{''})$, where
\begin{eqnarray}\nonumber
W^{'}(y_1,y_2|u_1) &=& \sum_{u_2}\frac{1}{2}W(y_1|u_1\oplus u_2)W(y_2|u_2) \\ \label{eq_wp}
 \\ \label{eq_wpp}
W^{''}(y_1,y_2,u_1|u_2) &=& \frac{1}{2}W(y_1|u_1\oplus u_2)W(y_2|u_2)
\end{eqnarray}
The Bhattacharyya parameters of channel $W^{'}$ and $W^{''}$ satisfy the following conditions:
\begin{eqnarray} \label{eq_zp}
Z(W^{''}) = Z(W)^2 \\ \label{eq_zpp1}
Z(W^{'}) \le 2Z(W)-Z(W)^2 \\ \label{eq_zpp2}
Z(W^{'}) \ge Z(W) \ge Z(W^{''})
\end{eqnarray}
The label $0$ (the upper branch in the transformation) in Fig.~\ref{fig_tree} means that the
output channel takes the branch $W^{'}$ in that specific transformation.
Correspondingly, a label $1$ (the lower branch in the transformation) means the output channel
takes $W^{''}$ in that transformation. Note that for binary
erasure channels (BEC), the Bhattacharyya parameter $Z(W^{'})$ has an exact expression $Z(W^{'}) = 2Z(W)-Z(W)^2$,
resulting in a recursive calculation of the Bhattacharyya parameters of the final bit channels.
Finally, after the channel transformations, the transition probability for bit channel $i$ is defined as

\begin{eqnarray}\label{eq_wn_slpit}
W_N^{(i)}(y_1^N,u_1^{i-1}|u_i) = \sum_{u_{i+1}^{N} \in \mathcal{X}^{N-i}}\frac{1}{2^{N-1}}W^N(y_1^N|u_1^NG)
\end{eqnarray}
where $W^N(\cdot)$ is the underlying vector channel ($N$ copies of the channel $W$) and $G$ is the generator
matrix whose form is to be discussed in the next section.

\subsection{Construction of Systematic Polar Codes}\label{sec_background_sys}
Polar codes in the original format \cite{arikan_iti09} are not systematic. The generator matrix for polar
codes is $G_p = B F^{\otimes n}$ in \cite{arikan_iti09} where $B$ is a permutation matrix and
$F=\left[\begin{smallmatrix} 1&0 \\ 1&1 \end{smallmatrix}\right]$. The operation $F^{\otimes n}$
is the $n$th Kronecker power of $F$ over the binary field $\mathbb{F}_2$. For systematic polar codes,
we focus on a generator matrix without the permutation matrix $B$, namely $G=F^{\otimes n}$. With such a matrix, the
encoding for normal polar codes is done as $\underline{x}=\underline{u}G$.

The indices of the source bits $\underline{u}$ corresponding
to the information bits can be set by selecting indices of the bit channels with the smallest Bhattacharyya parameters.
Denote $\mathcal{A}$ as the set consisting of indices for the information bits.
Correspondingly, $\bar{\mathcal{A}}$ consists of indices for the frozen bits.
Both sets $\mathcal{A}$ and $\bar{\mathcal{A}}$  are in $\{1, 2,..., N\}$.
For any element $i \in \mathcal{A}$ and $j \in \bar{\mathcal{A}}$, we have $Z(W_N^{(i)}) < Z(W_N^{(j)})$.
In this paper, the set $\mathcal{A}$ is always sorted in ascending order according to the index values,
instead of ordered by their values of Bhattacharyya parameters.

The source bits $\underline{u}$ can be
split as $\underline{u} = ({u}_{\mathcal{A}}, {u}_{\bar{\mathcal{A}}})$.
The codeword can then be expressed as
$\underline{x} = {u}_{\mathcal{A}}G_{\mathcal{A}}+ {u}_{\bar{\mathcal{A}}}G_{\bar{\mathcal{A}}}$,
where $G_{\mathcal{A}}$ is the submatrix of $G$ with rows specified by the set $\mathcal{A}$.
The systematic polar code is constructed by
specifying a set of indices of the codeword $\underline{x}$ as the indices to convey the information bits. Denote this
set as $\mathcal{B}$ and the complementary set as $\bar{\mathcal{B}}$. The codeword $\underline{x}$ is thus split as
$({x}_{\mathcal{B}},{x}_{\bar{\mathcal{B}}})$. With some manipulations, we have
\begin{equation} \label{eq_xb_xbc}
({x}_{\mathcal{B}},{x}_{\bar{\mathcal{B}}}) = ({u}_{\mathcal{A}}G_{\mathcal{AB}}+{u}_{\bar{\mathcal{A}}}G_{\bar{\mathcal{A}}\mathcal{B}}, {u}_{\mathcal{A}}G_{\mathcal{A\bar{B}}}+{u}_{\bar{\mathcal{A}}}G_{\mathcal{\bar{A}\bar{B}}})
\end{equation}
The matrix $G_{\mathcal{AB}}$ is a submatrix of the generator matrix with elements
$\{G_{i,j}\}_{i \in \mathcal{A}, j \in \mathcal{B}}$. Given a non-systematic encoder $(\mathcal{A},u_{\mathcal{\bar{A}}})$,
there is a systematic encoder $(\mathcal{B},u_{\mathcal{\bar{A}}})$ which performs the mapping
${x}_{\mathcal{B}} \mapsto \underline{x}=({x}_{\mathcal{B}},{x}_{\bar{\mathcal{B}}})$. To realize this systematic
mapping, ${x}_{\bar{\mathcal{B}}}$ needs to be computed for any given information bits ${x}_{\mathcal{B}}$. To this
end, we see from (\ref{eq_xb_xbc}) that ${x}_{\bar{\mathcal{B}}}$ can be computed if $u_{\mathcal{A}}$ is known.
The vector $u_{\mathcal{A}}$ can be obtained as the following
\begin{equation} \label{eq_ua}
u_{\mathcal{A}} = (x_{\mathcal{B}}-u_{\bar{\mathcal{A}}}G_{\mathcal{\bar{A}B}})(G_{\mathcal{AB}})^{-1}
\end{equation}
From (\ref{eq_ua}), it's seen that $x_{\mathcal{B}} \mapsto u_{\mathcal{A}}$ is one-to-one if $x_{\mathcal{B}}$
has the same elements as $u_{\mathcal{A}}$ and if $G_{\mathcal{AB}}$ is invertible.
In \cite{arikan_icl11}, it's shown that $\mathcal{B} = \mathcal{A}$ satisfies all these conditions in order to establish the
one-to-one mapping $x_{\mathcal{B}} \mapsto u_{\mathcal{A}}$. In the rest of the paper, the systematic encoding of polar
codes adopts this selection of $\mathcal{B}$ to be $\mathcal{B} = \mathcal{A}$. Therefore we can rewrite (\ref{eq_xb_xbc}) as
\begin{equation} \label{eq_xb_xbc_2}
({x}_{\mathcal{A}},{x}_{\bar{\mathcal{A}}}) = ({u}_{\mathcal{A}}G_{\mathcal{AA}}+{u}_{\bar{\mathcal{A}}}G_{\bar{\mathcal{A}}\mathcal{A}}, {u}_{\mathcal{A}}G_{\mathcal{A\bar{A}}}+{u}_{\bar{\mathcal{A}}}G_{\mathcal{\bar{A}\bar{A}}})
\end{equation}

\subsection{Theorem on Polar Coding Construction}
In this section, we prove a general theorem on polar codes. In the following, we say that
row $i$ intersects with column $j$ of the matrix $G$ if $G_{i,j} = 1$. Otherwise, we say row
$i$ does not intersect with column $j$.
\newtheorem{theorem}{Theorem}
\begin{theorem}\label{theorem_1}
For  $\forall j \in \mathcal{A}$ and $\forall i \in \mathcal{\bar{A}}$,
 row $i$ does not intersect with column $j$. Or in other words $G_{i,j} = 0$
 if $j \in \mathcal{A}$ and $i \in \mathcal{\bar{A}}$.
\end{theorem}
\begin{IEEEproof}
For any given index $j \in \mathcal{A}$, we divide the elements of $\mathcal{\bar{A}}$ into two sets:
$\bar{\mathcal{A}}_l = \{i: ~i \in \bar{\mathcal{A}}, ~ i< j\}$ and $\bar{\mathcal{A}}_g = \{i: i \in \bar{\mathcal{A}},~ i> j\}$.
For $i \in \bar{\mathcal{A}}_l$, it's obvious that $G_{i,j} = 0$ since the matrix $G$ is lower triangular. So we only need
to prove $G_{i,j} = 0$ for $i \in \bar{\mathcal{A}}_g$.

Let $(b_n^i,b_{n-1}^i,...,b_1^i)$ be the n-bit binary expansion of the integer $i-1$ with $i \in \bar{\mathcal{A}}_g$,
and $b_n^{i}$ is the MSB. The bit $b_n^i$ corresponds to the root channel selection in Fig.~\ref{fig_tree} and
$b_1^i$ corresponds to the last channel selection. Each bit in the binary
vector $(b_n^i,b_{n-1}^i,...,b_1^i)$ defines a channel selection of the corresponding level in the tree of Fig.~\ref{fig_tree}.
For example, bit $b_m^i$ ($m \in \{1,2,...,n\}$) determines bit channel $i$ at level $m$ takes the upper branch
or the lower branch.

Suppose row $i$ intersects with column $j \in \mathcal{A}$, equivalent to
$G_{i,j} = 1$. We know the entry of the generator matrix $G$ can be calculated as \cite{arikan_iti09}
\begin{equation}\label{eq_Gij}
G_{i,j} = \prod_{m=1}^{n}(1\oplus b_m^j \oplus b_m^j b_m^i)
\end{equation}
To have $G_{i,j} = 1$, we must have $b_m^i = 1$ when $b_m^j = 1$. Suppose $M_j$ is the last non-zero position
of $(b_n^j,b_{n-1}^j,...,b_1^j)$ and $M_i$ is the last non-zero position
of $(b_n^i,b_{n-1}^i,...,b_1^i)$. With $i \in \bar{\mathcal{A}}_g$, $M_i \ge M_j$. We proceed by discussing two
cases: $M_i = M_j$ and $M_i > M_j$.
\subsubsection{Case 1 $M_i = M_j$}
For $M_i = M_j$, we have
$(b_n^j,b_{n-1}^j,...,b_{M_j+1}^j) = (b_n^i,b_{n-1}^i,...,b_{M_j+1}^i) = 0_1^{n-M_j}$. Referring to Fig.~\ref{fig_tree},
it's seen that the recursive channel transformation from level $n$ to level $M_j+1$ (or $M_i+1$) is the same for both bit
channel $i$ and bit channel $j$:
they all take the upper branch in each transformation. Then at level $M_j$, both channels involve in the same fashion by taking
the lower branch (corresponding to $b_{M_j}^i = b_{M_j}^j =1$). Divide the levels $\{m: m \le M_j\}$ of bit channel $j$
into two sets:
\begin{eqnarray} \label{eq_m0}
\mathcal{M}_0 &=& \{m: m \le M_j ~~\text{and}~~ b_m^j = 0\}\\\label{eq_m1}
\mathcal{M}_1 &=& \{m: m \le M_j ~~\text{and}~~ b_m^j = 1\}
\end{eqnarray}
With $b_m^i = 1$ whenever $b_m^j = 1$, we can equivalently express $\mathcal{M}_1$  as
\begin{equation}\label{eq_m1_2}
\mathcal{M}_1 = \{m: m \le M_j ~~\text{and}~~ b_m^j = b_m^i=1\}
\end{equation}
Define a set $\mathcal{M}_{01} = \{m: m \in \mathcal{M}_0 ~~ \text{and}~~b_m^i = 1\}$. This
set $\mathcal{M}_{01}$ is not empty since
there must be at least one $m' \in \mathcal{M}_{0}$
at which $b_{m'}^i = 1$  since $i > j$, $M_i=M_j$, and
$b_m^i = 1$ whenever $b_m^j=1$ for $m \in \mathcal{M}_1$. When $|\mathcal{M}_{01}| > 1$, we select $m'$ to be the largest in
$\mathcal{M}_{01}$.
 At level $m'$, bit channel $i$ takes the lower branch
(corresponding to $b_{m'}^i = 1$)
and bit channel $j$ takes the upper branch (corresponding to $b_{m'}^j = 0$).
Therefore starting from level $m'$, the Bhattacharyya parameter for bit channel $i$ and bit channel $j$ diverge according
to (\ref{eq_zp}) and (\ref{eq_zpp2}):
\begin{eqnarray}
Z(W_{N_{m'}*2}^{(k_i)}) &=& (Z(W_{N_{m'}}^{(k_{m'})}))^2 \\
Z(W_{N_{m'}*2}^{(k_j)}) &\ge& Z(W_{N_{m'}}^{(k_{m'})})) \ge Z(W_{N_{m'}*2}^{(k_i)})
\end{eqnarray}
where
\begin{eqnarray}
N_{m'} &=& 2^{n-m'} \\
k_{m'} &=& (b_n,b_{n-1},...,b_{m'+1})\\
k_i &=& (b_n^i,b_{n-1}^i,...,b_{m'+1}^i,1) \\
k_j &=& (b_n^j,b_{n-1}^j,...,b_{m'+1}^j,0)
\end{eqnarray}
The number $k_i$ and $k_j$ is the channel index for bit channel $i$ and bit channel $j$ at level $m'$, respectively.
Starting from the same
previous channel $W_{N_{m'}}^{(k_{m'})}$, it's obvious that bit channel $i$ has a smaller Bhattacharyya parameter
than bit channel $j$ at level $m'$. For levels $m < m'$, this advantage of bit channel $i$ continues until
the last level because of the recursive channel transformation process defined by the set $\mathcal{M}_0$ and $\mathcal{M}_1$ in (\ref{eq_m0}) and (\ref{eq_m1_2}).
Therefore if $D_{i,j}=1$, when $M_i=M_j$, we have $Z(W_N^{(i)}) \le Z(W_N^{(j)})$
for $j \in \mathcal{A}$ and $i \in \mathcal{\bar{A}}$.

\subsubsection{Case 2 $M_i > M_j$}
In this case, we define a set $\mathcal{M}_{i1} = \{m: m>M_j ~~\text{and}~~ b_m^i = 1\}$. This set $\mathcal{M}_{i1}$ obviously
is not empty since $M_i > M_j$. But the set $M_{01}$ could be empty in this case. If $\mathcal{M}_{01} = \emptyset$, the recursive
channel transformation for bit channel $i$ and $j$ is the same for levels $\{m \le M_j\}$: they take the upper branches at levels
in $\mathcal{M}_0$ and take lower branches for levels in $\mathcal{M}_1$. However, their involving processes differ in at least
one level $m' \in \mathcal{M}_{i1}$ because of the existence of the non-empty set $\mathcal{M}_{i1}$.
When $|\mathcal{M}_{i1}| > 1$, we select
$m'$ to be the smallest in $\mathcal{M}_{i1}$.
Bit channel $i$ takes the lower branch at level
$m'$ while bit channel $j$ takes the upper branch at the same level, resulting in a smaller Bhattacharyya parameter for bit channel $i$ at that level. After level $m'$, as we already point out, the two channels involving in the same fashion defined by $\mathcal{M}_0$ and $\mathcal{M}_1$. Therefore the final Bhattacharyya parameters for bit channel $i$ is still smaller than bit channel $j$. If $\mathcal{M}_{01} \neq \emptyset$, the advantage of the bit channel $i$ is even more pronounced than the case when
$\mathcal{M}_{01} = \emptyset$ since bit channel $i$ takes additional lower branches besides taking the same lower branches as bit channel $j$, producing a final channel with an even smaller Bhattacharyya parameter. Therefore as in the case when $M_i = M_j$,  we also have $Z(W_N^{(i)}) \le Z(W_N^{(j)})$ for $j \in \mathcal{A}$ and $i \in \mathcal{\bar{A}}$ when $M_i > M_j$.

Combing Case 1 and Case 2, we see that if $D_{i,j} = 1$,  $Z(W_N^{(i)}) \le Z(W_N^{(j)})$ for $j \in \mathcal{A}$ and $i \in \mathcal{\bar{A}}$. But this contradicts with the polar encoding principle that
$Z(W_N^{(i)}) > Z(W_N^{(j)})$ for $j \in \mathcal{A}$ and $i \in \mathcal{\bar{A}}$. Therefore $D_{i,j} = 0$ for $j \in \mathcal{A}$ and $i \in \mathcal{\bar{A}}$.
\end{IEEEproof}

\newtheorem{corollary}{Corollary}
\begin{corollary}\label{corollary_gaca}
The matrix $G_{\mathcal{\bar{A}}\mathcal{A}} = 0$.
\end{corollary}

\begin{IEEEproof}
The statement of $G_{\mathcal{\bar{A}}\mathcal{A}} = 0$ is equivalent to say that any column $j \in \mathcal{A}$ of the
generator matrix $G$ does not intersect with row $i \in \mathcal{\bar{A}}$ of $G$, which we already prove in Theorem \ref{theorem_1}.
\end{IEEEproof}

Using Corollary \ref{corollary_gaca}, the systematic encoding of polar codes can be simplified as
\begin{equation} \label{eq_xb_xbc_3}
({x}_{\mathcal{A}},{x}_{\bar{\mathcal{A}}}) = ({u}_{\mathcal{A}}G_{\mathcal{AA}}, {u}_{\mathcal{A}}G_{\mathcal{A\bar{A}}}+{u}_{\bar{\mathcal{A}}}G_{\mathcal{\bar{A}\bar{A}}})
\end{equation}
The calculation of $u_{\mathcal{A}}$ in (\ref{eq_ua}) can thus be simplified as
$u_{\mathcal{A}} = x_{\mathcal{A}}G_{\mathcal{AA}}^{-1}$.

From the proof of Theorem \ref{theorem_1}, another corollary is readily available.
\begin{corollary}\label{corollary_intersect}
For any $i,j \in \mathcal{A}$, if row $i$ intersects with column $j$ of the generator matrix $G$ ($G_{i,j} = 1$),
then $Z(W_N^{(i)}) \le Z(W_N^{(j)})$.
Or in other words, bit channel $i$ has a better channel quality than bit channel $j$ when $G_{i,j} = 1$.
\end{corollary}

\subsection{Generator Matrix with Permutation}\label{sec_gp}
The original generator matrix in \cite{arikan_iti09} is $G_p = BF^{\otimes n}$ where $B$ is the bit-reversal permutation
matrix. We use the vector $\underline{a}$ to represent the sorted elements in $\mathcal{A}$ and the vector $\underline{b}$ the
corresponding vector consisting of the indices for the systematic encoding $\mathcal{B}$. In \cite{arikan_icl11}, it's pointed
out that $\mathcal{B}$ is the image of $\mathcal{A}$ under the matrix $B$, namely $\underline{b} = \underline{a}B$.
If the encoding of the normal polar codes is based on $G_p$, then the
submatrix $G_{\mathcal{AB}}$ in (\ref{eq_xb_xbc}) is $G_{\mathcal{AB}}= (G_p)_{\mathcal{AB}}$.
With some manipulations, it can be shown that $(G_p)_{\mathcal{AB}} = G_{\mathcal{AA}}$.
Thus, for systematic encoding, the generator matrix
$G_p = BF^{\otimes n}$ and $\underline{b} = \underline{a}B$ is equivalent to $G=F^{\otimes n}$ and $\mathcal{B} = \mathcal{A}$.
In the sequel, when it comes to the SC decoding, we assume the encoding is based
on the generator matrix $G_p$ so that the natural order schedule of the decoding can be applied.
This is only for the ease of description and doesn't affect the performance of systematic polar codes.

\section{The Theoretical Performance of Systematic Polar Codes} \label{sec_performance}
In this section, we provide a general relationship between the error performance of systematic polar codes
and the error performance of the non-systematic polar codes.

Denote the BER of non-systematic polar codes as $P_{b}$ and the corresponding BER of systematic polar codes as $P_{sys,b}$.
Define a set $\mathcal{A}_t \subseteq \mathcal{A}$ to contain the indices of the information bits in error for non-systematic
polar codes. Correspondingly,
the set $\mathcal{A}_{sys,t} \subseteq \mathcal{A}$ is the indices of the information bits in error for systematic polar codes under
the SC-EN decoding. The BER for systematic polar codes can be predicted from the BER of the
non-systematic polar codes in the following way:
\begin{equation} \label{eq_psys_b}
P_{sys,b} = \frac{\sum\limits_{\mathcal{A}_{sys,t}\subseteq \mathcal{A}}{|\mathcal{A}_{sys,t}|\Pr\{\mathcal{A}_{sys,t}\}}}{\sum\limits_{\mathcal{A}_t \subseteq \mathcal{A}}{|\mathcal{A}_{t}|\Pr\{\mathcal{A}_t\}}}P_b
\end{equation}
where $|\mathcal{A}_t|$ is to take the cardinality of the set $\mathcal{A}_t$ and $\Pr(\cdot)$ is the probability of the
inside event.

For any given set $\mathcal{A}_t$, the set $\mathcal{A}_{sys,t}$ can be calculated from it.
From (\ref{eq_xb_xbc_3}), we already have ${x}_{\mathcal{A}} = {u}_{\mathcal{A}}G_{\mathcal{AA}}$. This says that
the values or the errors in $x_{\mathcal{A}}$ only depend on ${u}_{\mathcal{A}}$ and $G_{\mathcal{AA}}$.
The values of the frozen bits don't affect the values or the errors of $x_{\mathcal{A}}$. Therefore, we can convert the
cardinality of the set $\mathcal{A}_t$ and $\mathcal{A}_{sys,t}$ into weight of the following vectors.
Let $\underline{v}$ be a $N$-element vector with $1$s
in the positions specified by $\mathcal{A}_t$ and $0$s elsewhere, namely $v_{\mathcal{A}_t}=\mathbf{1}_1^{|\mathcal{A}_t|}$.
Then the cardinality of the set $\mathcal{A}_t$ is the same
as the Hamming weight of the vector $\underline{v}$, written as $w_H(\underline{v})$. In the same way, we define a vector $\underline{q}$ with $q_{\mathcal{A}_{sys,t}}=\mathbf{1}_1^{|\mathcal{A}_{sys,t}|}$ and $0$s elsewhere.
We can then have
\begin{eqnarray}\label{eq_psys_b_1}
P_{sys,b} &=& \frac{\sum\limits_{\mathcal{A}_{sys,t} \subseteq \mathcal{A}}w_H(\underline{q})\Pr\{\mathcal{A}_{sys,t}\}}{\sum\limits_{\mathcal{A}_t \subseteq \mathcal{A}}w_H(\underline{v})\Pr\{\mathcal{A}_{t}\}}P_b \\\label{eq_psys_b_2}
&=& \frac{\sum\limits_{\mathcal{A}_t \subseteq \mathcal{A}}w_H(\underline{v}G)\Pr\{\mathcal{A}_{sys,t}\}}{\sum\limits_{\mathcal{A}_t \subseteq \mathcal{A}}w_H(\underline{v})\Pr\{\mathcal{A}_{t}\}}P_b
\end{eqnarray}
The equality $\underline{q} = \underline{v}G$ in equation (\ref{eq_psys_b_2})
is because of the re-encoding of $\underline{\hat{x}}=\underline{\hat{u}}G$ after the decoding of $\underline{\hat{u}}$. Note
that the operation $\underline{q} = \underline{v}G$ only represents the error
conversion from $\underline{v}$ to $\underline{q}$, not the real calculation of $\underline{\hat{x}}=\underline{\hat{u}}G$.

The cardinality of $\mathcal{A}$ is $|\mathcal{A}|=NR=K$ where $R$ is the code rate and $K$ is the number of information bits in each code block. It's easy to verify that the number of terms in the denominator of (\ref{eq_psys_b_2}) is $2^{NR}=2^K$.
With a large block length $N$ and a fixed code rate $R$,
it's  practically impossible to evaluate the error performance for systematic polar codes conditioned
on the error performance of the non-systematic polar codes.

In this section,
without considering the probabilities of the error events  $\{\mathcal{A}_t\}$,
we evaluate the error performance of systematic polar codes in two special cases to gain some initial insights of
the behavior of the systematic polar codes.
These two special cases are: 1) $v_{\mathcal{A}}=\mathbf{1}_1^K$;
and 2) The $e$th element of $\underline{v}$ is one: $v_e = 1$ with $e \in \mathcal{A}$. Case 1) is the situation where all bits are in error and case 2) says only one bit is in error.
The rationale for evaluating case 1) is due to the fact that if one bit $j \in \mathcal{A}$
is in error, then theoretically this error bit could affect all bits after it. This can be seen from the transition probability
of bit channel  $i > j$ in (\ref{eq_wn_slpit}): bit channel $i$ has its output $y_1^N$ (all received channel
samples) and $u_1^{i-1}$ (all previously decoded bits). As for case 2), it's related to the common assumption of coded systems that errors of the code bits in one codeword are independent and that at high SNR, there is only one bit in error in each codeword,
resulting in the relationship $P_b = P_s/N$, where $P_b$ is the BER and $P_s$ is the block error rate.

Before we analyze case 1, we need the following proposition.
\newtheorem{proposition}{Proposition}
\begin{proposition}\label{proposition_1}
For a block length $N=2^n$, $n \ge 0$, any column $j$ ($1 \leq j \leq N$) of the generator matrix $G=F^{\otimes n}$
has a Hamming weight of $2^{w_H(\bar{b}_1^j,\bar{b}_2^j,...,\bar{b}_n^j)}$,
where $(b_1^j,b_2^j,...,b_n^j)$ is the binary expansion of $j-1$ , and $\bar{b}_i^j = b_i^j \oplus 1$ over $\mathbb{F}_2$.
\end{proposition}

\begin{IEEEproof}
For a fixed column $j$, the weight of this column is to sum over all possible values of $i-1=(b_1^i,b_2^i,...,b_n^i)$: $\sum_{i}G_{i,j} = \sum_{i}F^{\otimes n}_{i,j}= \sum_{i}\prod_{m=1}^{n}(1\oplus b_m^j \oplus b_m^j b_m^i)$.
The rest of the proof is readily available.
\end{IEEEproof}

\subsection{All Bits in Error}\label{sec_all_error}
From Proposition \ref{proposition_1}, it can be inferred that except column $N$, the weight of all other columns of $G$ is even.
From Theorem \ref{theorem_1}, we know column $j \in \mathcal{A}$ of the generator matrix $G$ only has
$1$s at positions specified by
$\mathcal{A}$ since column $j$ doesn't
intersect with rows in $\mathcal{\bar{A}}$. Therefore, during the re-encoding process $\underline{q} = \underline{v}G$, the vector $q_{\{\mathcal{A}{\backslash} N\}}= \mathbf{0}_1^{N-1}$ and $q_N=1$. Here $\mathcal{A} \backslash N$ means the set $\mathcal{A}$ excluding the last element $N$.
The weight of $\underline{q}$ is then $w_H(\underline{q})=1$. We see almost all the errors in
the vector $\underline{v}$ are cancelled after the re-encoding process (with only one error remaining).
If this is the only error case, then $P_{sys,b} = \frac{1}{NR} P_b = \frac{1}{NR}$.
We give an example below to explicitly present this error cancelling process.

Suppose we are dealing with a BEC channel with an erasure probability $0.4$ and $N=16$.
Let the code rate $R = 1/2$. The code index set
can be calculated as $\mathcal{A} = \{8,10,11,12,13,14,15,16\}$. With all bits in error during the SC decoding process, the vector
$v_{\mathcal{A}} = \mathbf{1}_1^8$. The elements of $q_\mathcal{A}$ can be calculated from $q_{\mathcal{A}}=(\underline{v}G)_{\mathcal{A}}$.
For example,
\begin{eqnarray}
q_8 &=& v_8 + v_{16} = 0 \\
q_{10} &=& v_{10} + v_{12} + v_{14} + v_{16} = 0 \\
q_{11} &=& v_{11} + v_{12} + v_{15} + v_{16} = 0
\end{eqnarray}
With the weight of the columns of $G$ be even (excluding column $N$) and the columns with indices in $\mathcal{A}$ only intersect with rows in $\mathcal{A}$, the elements of $\underline{q}$ (excluding $q_N$) are essentially summing over even numbers of elements of $v_{\mathcal{A}}$, which eventually resulting in $0$s when $v_{\mathcal{A}} = \mathbf{1}_1^8$. The last element is $q_{16} = v_{16} = 1$, which is the only error remaining after the vector $\underline{v}$ going through the matrix $G$.
The error rate is then $P_b=1$ and $P_{sys,b} = \frac{1}{8}$.

From this example, it's seen that the re-encoding process of $\underline{\hat{x}}=\underline{\hat{u}}G$ after decoding
$\underline{\hat{u}}$ does not amplify the number of errors in $\underline{\hat{u}}$ when all bits of ${\hat{u}}_{\mathcal{A}}$
are in error. Actually in this case, the number of errors is already at its maximum and can't be amplified.
But the number of errors doesn't stay the same, as one would expect in this case, after the re-encoding process.
Instead, almost all errors are cancelled after the re-encoding process.

\subsection{One Bit in Error} \label{sec_one_error}
Now we return to the case with only one error $v_e=1$, $e \in \mathcal{A}$. Denote the $e$th row of $G$ as $G_{e,:}$.
The indices of the corresponding error bits for systematic polar codes are the indices of the non-zero positions of the
subvector $(G_{e,:})_\mathcal{A}$.
Therefore the number of non-zero positions of $q_{\mathcal{A}}$ is determined by the weight of this subvector $(G_{e,:})_\mathcal{A}$: $w_H(q) = w_H\{(G_{e,:})_\mathcal{A}\}$. Due to the fact that $G$ is a lower triangular matrix, only elements in $\{i: i \in \mathcal{A} ~~\text{and}~~ i \le e \}$ of ${q}_{\mathcal{A}}$ are affected by
this error in $\underline{v}$. In this one-error case, the number of errors could be amplified after the re-encoding process $\underline{\hat{x}}=\underline{\hat{u}}G$, depending on the location of the error.
The error rate is $P_b = \frac{1}{NR}$ and
$P_{sys,b} = \frac{w_H\{(G_{e,:})_\mathcal{A}\}}{NR} P_b$.
In the preceding example, instead of
$v_{\mathcal{A}} = \mathbf{1}_1^8$, if we only have $v_{16} = 1$, then $q_{\mathcal{A}} = \mathbf{1}_1^8$ since $w_H(G_{16,:})_{\mathcal{A}} = 8$,  resulting in $P_b=\frac{1}{8}$ and $P_{sys,b} = 1$.
But if we only have $v_8 = 1$
or $v_{10} = 1$, then we also only have the corresponding bit in error
 $q_8=1$ or $q_{10}=1$ with $P_b = P_{sys,b} = \frac{1}{8}$.
 The number of errors in the case $v_{16}=1$ is
indeed amplified by $8$ times after the re-encoding process while the number of errors with $v_8 = 1$
or $v_{10} = 1$ stays the same after the re-encoding process.

\section{Systematic Polar Codes Gain} \label{sec_error_pattern}

In the discussions from Section \ref{sec_all_error} and \ref{sec_one_error}, we already see that the number of
errors of polar codes with the SC decoding is not necessarily amplified in the SC-EN decoding process of
systematic polar codes.
It all depends on how the errors are distributed in the SC decoding process.
This section is devoted to the analysis of the behavior of the errors in the SC decoding process and to
characterize the advantage of systematic polar codes over non-systematic polar codes.
The analysis is based on BEC channels. In
Section \ref{sec_numerical}, it's seen that the results in this Section can be extended to AWGN channels as well.

\subsection{Basic Error Patterns}\label{sec_basic_error_pattern}

In order to understand how the errors are distributed with the SC decoding, we first look at the basic error patterns.
The decoding graph of polar codes with a block length $N=2^n$ consists of $n$ columns of Z-shape sections,
with each column having
$N/2$ Z-shape sections. For the connections of the Z-shape sections in each level, please refer to \cite{arikan_iti09}\cite{urbanke_isit09}.
In this subsection, we use the natural order schedule for the SC decoding as discussed in Section \ref{sec_gp}.

\begin{figure*}
{\par\centering
\resizebox*{3.0in}{!}{\includegraphics{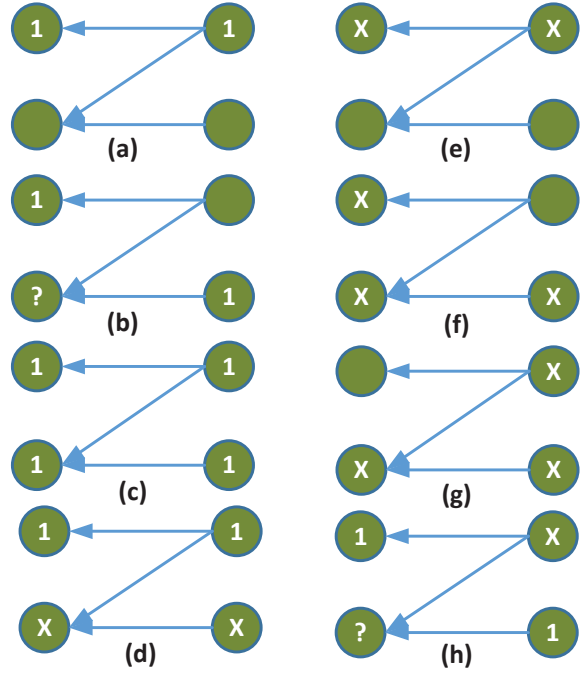}} \par}
\caption{Basic Error Patterns. A variable node without a label means its LR value is correct. The meanings of the labels are:
X referring to an incorrect LR;
1 meaning a LR value one; and label ? referring to a LR value which could be correct or incorrect.}
\label{fig_error_pattern}
\end{figure*}

The basic error patterns in the decoding graph are illustrated in
Fig.~\ref{fig_error_pattern}, where a node without any label has a correct likelihood ratio (LR) value, a node with a
label 1 has a LR value of one,
a node with a label X has an incorrect LR value, and a node with a label ? can have a correct or incorrect LR value depending
on the context. In the SC decoding, before the first error happens, the LR
values of the variable nodes in the Z-shape sections are either correct or 1, represented
by (a),(b) and (c) of Fig.~\ref{fig_error_pattern}. We provide the proof of the error pattern
Fig.~\ref{fig_error_pattern}-e in the Appendix and
all other patterns in Fig.~\ref{fig_error_pattern} can be proved in the same fashion.

The LR value of the first error bit must be one. Again, the proof of this fact is omitted as this is relatively a simple
practice. In other words, the first error happens because the decoder takes an incorrect guess,
corresponding to the upper left node in Fig.~\ref{fig_error_pattern}-(a)(b)(c) and the lower left node in Fig.~\ref{fig_error_pattern}-(c).

After the first error, as we already point out, all bits after this error bit could potentially be affected by this error.
For example, the lower left nodes in Fig.~\ref{fig_error_pattern}-(d)(f)(g) are in error because of the previous errors.
These errors are surely the errors propagated (or coupled) from previous errors.
But not all bits after the first error bit are in error,
simply through observing the basic error patterns in Fig.~\ref{fig_error_pattern}.
The first example is Fig.~\ref{fig_error_pattern}-(a). If the bit (or the combined bit) corresponding to the upper
left node is in error, then the LR value corresponding to the lower left bit is still correct. Actually, the LR of the lower
left node is not affected by the upper left node since the upper right node in Fig.~\ref{fig_error_pattern}-(a)
has a LR value of one.
In this case, as long as the lower right node has a correct LR value, the lower left node can always make a correct decision.
Another example is Fig.~\ref{fig_error_pattern}-(e) in which the upper left node has an incorrect LR
value thus with an incorrect bit decision.
But the incorrect bit decision cancels the effect of the incorrect LR value of the upper right node when it comes
to the decision of the lower left node. Therefore the lower left node can make a correct decision in this case even though the
upper left node has an incorrect decision. For a rigorous proof of this pattern, please refer
to the Appendix. There are other cases, for example Fig.~\ref{fig_error_pattern}-(g)(h),
where incorrect LRs due to incorrect previously decoded bits
don't necessarily cause all bits in error after those error bits.
Because of these effects, it's extremely unlikely that after the first error bit,
all bits after it are in error, especially with large block lengths. For the same reason,
it's also unlikely that all bits after the
first error bits are correct. In other words, one bit error, like all bits in error, is also unlikely.

From the basic error patterns in Fig.~\ref{fig_error_pattern}, one proposition can be easily obtained for BEC channels.
\begin{proposition}
For polar codes with the SC decoding on BEC channels, the number of nodes with $LR = 1$
stays the same in each column of the decoding graph.
\end{proposition}

\subsection{First Error Distribution}
As stated in the previous section, the first error happens because the decoder takes
an incorrect guess. All calculations before
the first error involve patterns in Fig.~\ref{fig_error_pattern}-(a)(b)(c). Note that the question marker in the lower left node should be removed before the first error as there are no errors yet. Of course, there is always a pattern involving two correct
nodes which is not shown in Fig.~\ref{fig_error_pattern}.

The probability of bit $i$ being the first error is determined by the quality of bit channel $i$, which in turn is determined by its Bhattacharyya parameter. For a rigorous proof, please refer to Section V-B of \cite{arikan_iti09}.
In this section, we present simulation results on the first error distribution without further theoretical discussions.

\begin{figure*}
{\par\centering
\resizebox*{3.0in}{!}{\includegraphics{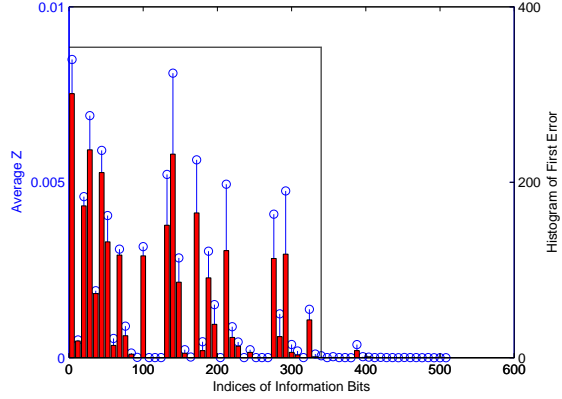}} \par}
\caption{First error histogram and the corresponding average Bhattacharyya Parameter. The code block length is $N=2^{10}$ and
the code rate is $R=1/2$. The underlying channel is the BEC channel with an erasure probability 0.4.
The right y-axis is for the bar plot and the left y-axis is for the stem plot. The labels of the x-axis are the indices of elements in sorted $\mathcal{A}$, not the real values of elements in $\mathcal{A}$.}
\label{fig_first_error}
\end{figure*}

For BEC channels, we can precisely calculate the Bhattacharyya parameter for each bit channel using the recursive
expressions given in \cite{arikan_iti09}. Fig.~\ref{fig_first_error} shows the histogram
of the indices of the first error bit
and the corresponding average Bhattacharyya parameter for $N=2^{10}$ and $R=1/2$ in a BEC channel with
an erasure probability $0.4$.
Fig.~\ref{fig_first_error} has two
y-axes: the right axis shows the number of occurrences of the first error and the left axis shows the value of the corresponding Bhattacharyya parameters. Seen from Fig.~\ref{fig_first_error}, the probability of the first error is indeed determined by the quality of each bit channel. Also shown in Fig.~\ref{fig_first_error} is the brick-wall nature of the first
error distribution, which is the reflection of the polarization effect of the $N$ channels.

At this point, we want to point out the effect of the first error in non-systematic and systematic polar codes scenarios.
The first error in the SC decoding process could potentially affect
all bits after it (or bits with indices larger than it with the natural order decoding).
This effect can be considered as a forward error effect.
But in the re-encoding process of the SC-EN decoding of systematic polar codes,
the errors (including the first error) in the decoded
vector $\hat{\underline{u}}$ only affect bits in $\hat{\underline{x}}$ before them (or bits with indices smaller than the error bits), due to the lower triangularity of the generator matrix. Correspondingly, this effect
in the re-encoding process is a backward error effect.


\subsection{Systematic Polar Codes Gain}\label{sec_error_model}
In this section, we extract the first part of the right hand side of (\ref{eq_psys_b}) and define the reverse of it as
the gain of systematic polar codes over
non-systematic polar codes:
\begin{equation} \label{eq_gamma_e}
\gamma = \frac{\sum\limits_{\mathcal{A}_t \subseteq \mathcal{A}}{|\mathcal{A}_{t}|\Pr\{\mathcal{A}_t\}}}{\sum\limits_{\mathcal{A}_{sys,t}\subseteq \mathcal{A}}{|\mathcal{A}_{sys,t}|\Pr\{\mathcal{A}_{sys,t}\}}}
\end{equation}
From the previous discussion in Section \ref{sec_all_error} and \ref{sec_one_error}, we can safely constrain
the systematic gain to be strict for large $N$: $\frac{1}{NR} < \gamma < NR$.

The analysis in Section \ref{sec_performance} does not include the effect of the coupling between errors
as discussed in Section \ref{sec_error_pattern}. From the discussions in Section \ref{sec_basic_error_pattern},
we know the errors in previous decoded bits could affect the bits after them, although not all bits after the
error bits are necessarily in error. With a large $N$, there are $2^N-1$ error combinations in the received
vector $y_1^N$. Therefore, the errors of the decoded bits after the first decoded error
can be considered as independent and identically distributed (i.i.d) with probability $p$ when $N$ is large.
Based on this assumption, we can convert the calculation of the systematic gain in (\ref{eq_gamma_e})
into the analysis of a function involving the first error distribution and the probability $p$.

Denote $p_i$ as the probability of the first error occurring to the information bit $i$ and
denote this error event as $\xi_i$. As in
Section \ref{sec_performance}, we use a vector $v_1^N$ to represent the error positions: $v_i = 1$ if bit $i$ is in
error and $v_i = 0$ otherwise.
Then the probability of all information bits in error conditioned on $\xi_i$ is:
\begin{equation}\label{eq_prob_A}
\Pr\{v_\mathcal{A} = 1_1^K |\xi_i\} = (0, 0, ..., 1, p, p,...,p)
\end{equation}
In (\ref{eq_prob_A}), the first $(i-1)$ probabilities are zeros  because the first error at bit $i$ doesn't affect
bits before it (the forward error effect). After bit $i$, the errors
are i.i.d with probability $p$ as discussed previously. Since the events
$\{\xi_i\}_{i=1}^K$ are exclusive, the probability of the information bits in error is simply the following summation
\begin{equation}
\Pr\{v_\mathcal{A} = 1_1^K\} = \sum_{i=1}^K \Pr\{v_\mathcal{A} = 1_1^K |\xi_i\}
\end{equation}
with the individual bit error as $\Pr\{v_i=1\}= p_i + p\sum_{j=1}^{i-1}p_j$.

Utilizing the brick-wall property of the first error distribution as shown in
Fig.~\ref{fig_first_error}, we can divide the bits in
$\mathcal{A}$ into two groups: group one consisting of the error bits
due to the bad bit channel conditions and group
two consisting of the error bits purely coupled from group one.
Denote these two groups as $\mathcal{A}_I$ and $\mathcal{A}_C$ respectively. Referring to Fig.~\ref{fig_first_error},
the set $\mathcal{A}_I$ includes the bits within the brick wall and the set $\mathcal{A_C}$ includes the bits
outside the brick wall. Denote $K_I = |\mathcal{A}_I|$. The probabilities can now
be expressed as:
\begin{equation}\label{eq_vi_1}
\Pr\{v_i=1\} =
    \begin{dcases}
        p_i + p\sum_{j=1}^{i-1}p_j, ~ 1 \leq i \leq K_I \\
        p\sum_{j=1}^{K_I}p_j, ~~~~~~~ K_I  < i \leq K
    \end{dcases}
\end{equation}
And the systematic gain can be calculated using (\ref{eq_vi_1})  as
\begin{equation}\label{eq_gamma_2}
\gamma = \frac{\E\{{\omega_H(v_1^N)}\}}{\E\{{\omega_H(v_1^NG)}\}}
\end{equation}

The evaluation of (\ref{eq_gamma_2}) involves the distribution of the first error probabilities
$\{p_i\}_{i=1}^K$ of the information bits and the probability $p$. The distribution of the probabilities $\{p_i\}_{i=1}^K$
can be approximated by the distribution of the corresponding Bhattacharyya parameters
$\{Z(W_N^{(i)})\}$. But the combined effect of the probability $p$ and the distribution of $\{Z(W_N^{(i)})\}$ is not intended
to be fully discussed in this paper due to the space limit. Instead,
in Section \ref{sec_model}, we establish a simplified statistic model to
characterize the probabilities in (\ref{eq_vi_1}) and this model is
used to calculate the systematic gain $\gamma$ in (\ref{eq_gamma_2}).

\subsection{A Qualitative View of the Systematic Gain}

Using Corollary \ref{corollary_intersect}, we can qualify why the systematic gain $\gamma$ should be generally larger than one.
Or at least, the systematic polar codes should perform as well as the non-systematic polar codes.
In the re-encoding process, the estimation $\hat{\underline{x}} = \hat{\underline{u}}G$ is performed.
The entry of $\hat{x}_{\mathcal{A}}$, say $\hat{x}_j$ ($j \in \mathcal{A}$), is
\begin{equation}\label{eq_xi}
\hat{x}_j = \hat{u}_{\mathcal{A}}G_{\mathcal{A},j}
\end{equation}
where $G_{\mathcal{A},j}$ is the $j$th column of $G$ with entries specified by $\mathcal{A}$.
The error correction capability of the systematic
polar codes comes from this re-encoding process in (\ref{eq_xi}). To understand this capability of systematic
polar codes, we first note that the weight of all the columns of the matrix $G$ is even except the last column.
This property of $G$ is already stated in the beginning of Section \ref{sec_all_error}. This is where the theoretical
maximum $\gamma = NR$ comes from.

From (\ref{eq_xi}), it's seen that the errors in
$\hat{u}_{\mathcal{A}}$ can only affect $\hat{x}_j$ at positions where $G_{\mathcal{A},j}$ have non-zero
entries. From Corollary \ref{corollary_intersect}, it's known that a
non-zero entry of column $j$, $G_{i,j}=1$, means
a better bit channel $i$ than $j$. Let's call the set of bits $\{i: i \neq j, ~i \in \mathcal{A} ~\text{and}~ G_{i,j}=1\}$
the compatible bits of the information bit $j$.
For bit $\hat{x}_j$, only bit $j$ and its compatible bits affect the decision.
Since the compatible bits of bit $j$ transmit at better bit channels than $j$, it's more likely that bit $j$ is in error
and the compatible bits are in error due to the error propagation of bit $j$. In other words, the errors of bit $j$ and
its compatible bits are coupled. The re-encoding process $\hat{x}_j = \hat{u}_{\mathcal{A}}G_{\mathcal{A},j}$ is
equivalent to sum over bit $j$ and its compatible bits, a process to average out the coupled errors.
This mechanism of the re-encoding process leads to the fact that systematic polar codes perform at least
as well as non-systematic polar codes, or $\gamma \ge 1$.

\section{Composite Error Model}\label{sec_model}
So far we are still short of an efficient way to calculate the systematic gain $\gamma$. In this section, we establish
a statistic model to simplify the probabilities of the errors in (\ref{eq_vi_1}).
This simplified model is then used to calculate the systematic gain $\gamma$.

We define a new set $\mathcal{S}$ as the ensemble of the error events $\mathcal{A}_t$:
\begin{equation}\label{eq_set_s}
\mathcal{S} = \bigcup_{\substack{\mathcal{A}_t \subseteq \mathcal{A}}} \mathcal{A}_t
\end{equation}
Considering the basic error patterns in Fig.~\ref{fig_error_pattern}, the errors could happen to any bit after the first
error bit, no matter which bit channel the bit experiences.
Therefore, the set $\mathcal{S}$ can almost surely consist of all the information bits after the
first information bit with a non-negligible Bhattacharyya parameter.
For this, we set a threshold $\alpha$, below which the
Bhattacharyya parameter is considered as negligible. Otherwise, the Bhattacharyya parameter is considered as large.
Define a set consisting of all the indices of the bit channels with non-negligible Bhattacharyya parameters as
\begin{equation}\label{eq_large_z}
\mathcal{I} = \{i, ~~i \in \mathcal{A} ~\text{and}~ Z(W_N^{(i)}) > \alpha\}
\end{equation}
As stated in Section \ref{sec_background_sys}, the set $\mathcal{A}$ is sorted in ascending order
according to the index values. So is the set $\mathcal{I}$. This set $\mathcal{I}$ is used to define
the boundaries of the brick wall in Fig.~\ref{fig_first_error}.
The first element (with the smallest index value)
in $\mathcal{I}$ is denoted as $I_1$.
Then $\mathcal{S}$ can be  written as
\begin{equation}
\mathcal{S} = \{a: ~~a \in \mathcal{A} ~\text{and}~~ a \ge I_1\}
\end{equation}
When $I_1$ happens to be also the first element of $\mathcal{A}$, then $\mathcal{S} = \mathcal{A}$.
The elements of $\mathcal{S}$ are also sorted according to the index values as the elements in the set $\mathcal{A}$.

The next part to define $\mathcal{S}$ is to assign each element in $\mathcal{S}$ a probability of being in error.
Following the discussions in Section \ref{sec_error_model}, we perform the following steps to $\mathcal{S}$:
\begin{itemize}
\item Divide the set $\mathcal{S}$ into two sections:
the first section, denoted as $\mathcal{S}_{1}$, being the
region where the first error could happen, and the second section, $\mathcal{S}_{2}$,
being the region where errors are coupled or induced from
region one.
\item The composite effect of $\cup\mathcal{A}_t$ in the first region $\mathcal{S}_{1}$ is denoted by the error
probability of the first equation of (\ref{eq_vi_1}).
\item The composite effect of $\cup\mathcal{A}_t$ in the second region $\mathcal{S}_{2}$ is denoted by the error
probability of the second equation of (\ref{eq_vi_1}).
\end{itemize}

From the second equation of \ref{eq_vi_1},
it's known that all bits in $\mathcal{S}_2$ have the same probability of error from an composite point
of view and this probability of error should be larger than the
probability of error in $\mathcal{S}_1$  due to the following observation.
In region one, any bit with index $a_1$ could be in error at one error event
$\mathcal{A}_1 \subseteq \mathcal{A}$ with
 the first error bit $e_1 < a_1$, but will be for sure correctly decoded in another error event
$\mathcal{A}_2 \subseteq \mathcal{A}$ when $a_1 < e_2$ with $e_2$ being the index of the first error bit in event $\mathcal{A}_2$.
In region two, any bit can be potentially decoded incorrectly in any error event.
So statistically, the bits in region  $\mathcal{S}_{2}$ have a higher probability of being in error when considering the
composite effects of $\cup\mathcal{A}_t$. This condition translates to the way we
select the probability $p$ in (\ref{eq_vi_1}). However, as we point out in Section \ref{sec_error_model},
it's theoretically difficult to precisely calculate the probability of
error for each element in $\mathcal{S}$. With the above observation, we propose the following
simplified model in place of the precise model:
\begin{eqnarray}\nonumber
\mathcal{S}_{1} &=& \{a: ~a \in \mathcal{S}, ~a \le I_m, \\ \label{eq_st1}
&&~~~~\Pr\{\text{a is in error}\} = p_0  \} \\ \nonumber
\mathcal{S}_{2} &=& \{a: ~a \in \mathcal{S}, ~a > I_m, \\ \label{eq_st2}
&&~~~~\Pr\{\text{a is in error}\} = 1 \}
\end{eqnarray}
where $I_m$ is the last element in $\mathcal{I}$.
What this model says is the following: The mean effect of $\cup\mathcal{A}_t$ is that for information bits with indices
in $\mathcal{S}_{1}$, their errors are statistically independent with a probability $p_0$.
The rest of the information bits are in error with probability one from a composite point of view.
Although in this paper
a precise probability $p_0$ is not pursued, empirically we find that $p_0=1/2$ is a very good approximation.

An important parameter of the model in (\ref{eq_st1})(\ref{eq_st2}) is the boundary $I_m$.
It's clear that this boundary element
$I_m$ is related to the channel $W$. For example, with a BEC channel, when the block length $N$ and the code rate $R$ is fixed,
the boundary $I_m$ is related to the erasure probability. With a large erasure probability, there are more bits which are in error
due to the channel itself and less bits in error due to the forward error effect, and vice versa. Without going into the details
of calculating the Bhattacharyya parameters of the bit channels (which is only possible for BEC channels),
we can use a coupling coefficient to calculate another boundary element $\tilde{I}_m \in \mathcal{S}$.
The coupling coefficient here means the fraction of incorrect information
bits due to the previously incorrectly decoded information bits. Denote the coupling coefficient as $\beta$ and the
element $\tilde{I}_m$ is the $m'$th element of $\mathcal{S}$ where
\begin{eqnarray}\label{eq_m}
m'=\lfloor |\mathcal{S}|*(1-\beta)\rfloor
\end{eqnarray}
Then we can use $\tilde{I}_m$ to replace the boundary element $I_m$ in the model (\ref{eq_st1})(\ref{eq_st2}). This boundary
based on the coupling
coefficient is especially useful for bit channels whose Bhattacharyya parameters are not readily available.

Note that this simple model in (\ref{eq_st1})(\ref{eq_st2}) can approximate the composite effect
$\cup\mathcal{A}_t$ only in the statistical sense and it only models the dominant effect (or the mean effect) of $\cup\mathcal{A}_t$.
It is not, by any means, an exact error event $\mathcal{A}_t \subseteq \mathcal{A}$.

\subsection{Calculation of the Systematic Gain}
With the composite error model in (\ref{eq_st1})(\ref{eq_st2}),
we can calculate the systematic gain.
Use the same $N$-element vector $\underline{v}$ as an error indicator vector of $\mathcal{S}$:
the $i$th entry of $\underline{v}$ is zero if $i \notin \mathcal{S}$; otherwise $v_i$ is one
if $i \in \mathcal{S}$ and the $i$th bit is in error.
The subvector corresponding to region two of $\mathcal{S}$ is
$v_{\mathcal{S}_{2}}=\mathbf{1}_1^{|\mathcal{S}_{2}|}$ seen from (\ref{eq_st2}).
Each element of the subvector $v_{\mathcal{S}_{1}}$
takes value in $\{0,1\}$ with probability $p_0$ as shown in (\ref{eq_st1}).
The systematic gain from the composite model is thus
\begin{equation}\label{eq_gamma_c}
\gamma = \frac{\E\{w_H(\underline{v})\}}{\E\{w_H(\underline{v}G)\}}
\end{equation}
The mean weight of $\underline{v}$ can be easily calculated as
$\E\{\omega_H(\underline{v})\} = p_0|\mathcal{S}_1| + |\mathcal{S}_2|$.
Now we need to calculate the mean weight of $x_{\mathcal{S}} = \underline{v}G_{\mathcal{S}\mathcal{S}}$,
which can be decomposed as
$(x_{\mathcal{S}_{1}}, x_{\mathcal{S}_{2}}) = \underline{v}(G_{\mathcal{S}\mathcal{S}_{1}}, G_{\mathcal{S}\mathcal{S}_{2}})$. Due to the lower triangularity of the matrix $G_{\mathcal{S}\mathcal{S}}$,
the weight of $x_{\mathcal{S}_{2}}$ can be directly calculated as
\begin{equation}
\omega_H\{x_{\mathcal{S}_{2}}\} = \omega_H\{\underline{v}G_{\mathcal{S}\mathcal{S}_{2}}\} = \omega_H\{v_{\mathcal{S}_{2}}G_{\mathcal{S}_{2}\mathcal{S}_{2}}\} = 1
\end{equation}
which uses the even weight property of the columns of $G$ except the last column.
The first part $x_{\mathcal{S}_{1}} = \underline{v}G_{\mathcal{S}\mathcal{S}_{1}}$ can be
further divided into the summation of two parts:
\begin{equation}\label{eq_xst1}
x_{\mathcal{S}_{1}} = v_{\mathcal{S}_{1}}G_{\mathcal{S}_{1}\mathcal{S}_{1}} + v_{\mathcal{S}_{2}}G_{\mathcal{S}_{2}\mathcal{S}_{1}}
\end{equation}
The second part in (\ref{eq_xst1}) is a deterministic vector since $v_{\mathcal{S}_{2}}$ is the all-one
vector. With $G_{\mathcal{S}_{1}\mathcal{S}_{1}}$ an invertible lower triangular matrix, the vector
$x_{\mathcal{S}_{1}}$ belongs to the row space of the matrix $G_{\mathcal{S}_{1}\mathcal{S}_{1}}$.
Thus it can be formed by another vector $\tilde{v}_{\mathcal{S}_{1}}$ in the identity basis of the row space of
$G_{\mathcal{S}_{1}\mathcal{S}_{1}}$ as:
\begin{equation}
x_{\mathcal{S}_{1}} = \tilde{v}_{\mathcal{S}_{1}}I
\end{equation}
with $\tilde{v}_{\mathcal{S}_{1}}$ defined in the same way as ${v}_{\mathcal{S}_{1}}$.
Therefore the mean weight of $x_{\mathcal{S}_{1}}$ is the same as the mean weight of $\tilde{v}_{\mathcal{S}_{1}}$ which is
\begin{equation}
\E\{\omega_H\{x_{\mathcal{S}_{1}}\}\} = p_0|\mathcal{S}_{1}|
\end{equation}
The systematic gain is then
\begin{equation}\label{eq_mean_gamma}
\gamma = \frac{p_0|\mathcal{S}_{1}| + |\mathcal{S}_{2}|}{1+p_0|\mathcal{S}_{1}|}
\end{equation}
When the cardinality of $\mathcal{S}_{1}$ is quite large, the systematic gain can be approximated
as:
\begin{equation}\label{eq_mean_gamma_2}
\gamma \approx 1 + \frac{1}{p_0}\frac{|\mathcal{S}_{2}|}{|\mathcal{S}_{1}|}
\end{equation}
An immediate conclusion from (\ref{eq_mean_gamma_2}) is that the systematic gain is greater than one, meaning
that systematic polar codes should perform better than the corresponding non-systematic polar codes.
Another interpretation of (\ref{eq_mean_gamma_2}) is that the systematic gain is only determined by
the ratio of cardinalities of
the two sets $\mathcal{S}_1$ and $\mathcal{S}_{2}$. It does not increase with the increase of the block length as one would
intuitively expect. This property of the systematic polar codes is verified in the
simulations in Section \ref{sec_numerical}.


\section{Numerical Results} \label{sec_numerical}
In this sections, numerical examples for both BEC channels and AWGN channels are provided to validate
the results in Sections \ref{sec_error_pattern} and \ref{sec_model}. The encoding for BEC channels are
done through the selection of the bit channels with the smallest Bhattacharyya parameters. For AWGN
channels, we still use the same recursive formula in calculating the Bhattacharyya parameters for BEC
channels in encoding. We emphasize that this encoding serves our purpose just as well,
as long as it's consistent for both non-systematic polar codes and systematic polar codes.

\begin{figure*}
{\par\centering
\resizebox*{3.0in}{!}{\includegraphics{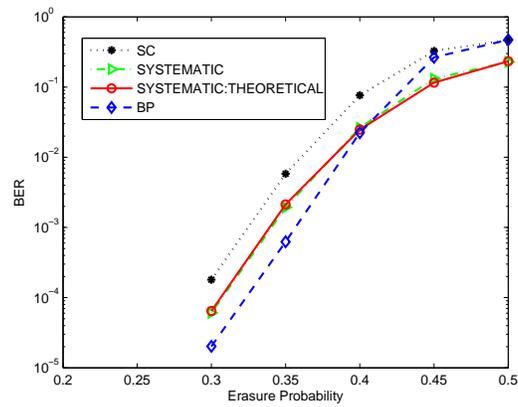}} \par}
\caption{BER for $n=10$, $R=1/2$ in BEC channel. }
\label{fig_ber_bec_n_10}
\end{figure*}

Fig.~\ref{fig_ber_bec_n_10} is the result in the BEC channel for $N=2^{10}$ and $R=1/2$.
Several curves are shown in Fig.~\ref{fig_ber_bec_n_10}. The curve of the stared dotted line
is the BER of the non-systematic polar codes under the SC decoding. The legend for this
curve is `SC'.
The curve of the dash dotted line with triangles is the BER of
the systematic polar codes with the SC-EN decoding for which the legend is `SYSTEMATIC'.
The circled solid line is the theoretical BER for systematic polar codes from the
model in (\ref{eq_st1})(\ref{eq_st2}).  Also shown in Fig.~\ref{fig_ber_bec_n_10} is the BER
of the non-systematic polar codes
with the belief-propagation (BP) decoding (the curve of the dashed line with diamonds).

The theoretical BER for systematic polar codes in Fig.~\ref{fig_ber_bec_n_10} is generated
using two different coupling coefficients: $\beta = 0.3$ when the erasure probability is
larger than 0.45 and $\beta = 0.5$  when the erasure probability is smaller than 0.45. This
choice of the coupling coefficient corresponds to the bad channel condition and the good channel
condition, respectively. The probability of independent error in $\mathcal{S}_{1}$
is $p_0=1/2$ and it's used in all of the following theoretical calculations.
 The threshold $\alpha$ in determining the set $\mathcal{I}$ in
(\ref{eq_large_z}) is set to be $\alpha = 10^{-3}$. Under this setting, the first element
of $\mathcal{I}$ is $I_1 = 192$ when the erasure probability is 0.4, which is also the first element of $\mathcal{A}$.
Thus the composite set is $\mathcal{S} = \mathcal{A}$ in this case. The systematic gain calculated from the composite
set $\mathcal{S}$ is quite stable. A small number can be used in averaging this systematic gain.
In Fig.~\ref{fig_ber_bec_n_10}, only ten realizations are used in
calculating the theoretical systematic gain $\gamma$.
The simulated BER and the BER from the model in (\ref{eq_st1})(\ref{eq_st2})
match quite well, showing that the simple model in (\ref{eq_st1})(\ref{eq_st2}) can
approximate the dominant error events of $\cup\mathcal{A}_t$ and thus can be used
to calculate the systematic gain.

Also showing in Fig.~\ref{fig_ber_bec_n_10} is the BER for non-systematic polar codes with the BP decoding.
BP decoding is generally better than the SC decoding as shown in \cite{urbanke_isit09}. With a bad
channel condition, for example, with an erasure probability larger than 0.45,
BP decoding performs almost the same as the SC decoding.
Systematic polar codes, however, perform two to three times
better than both SC and BP decoding under the same channel conditions,
at a cost almost negligible compared to the complexity of the BP decoding.
At better channel conditions, BP decoding starts to show its advantage.

\begin{figure*}
{\par\centering
\resizebox*{3.0in}{!}{\includegraphics{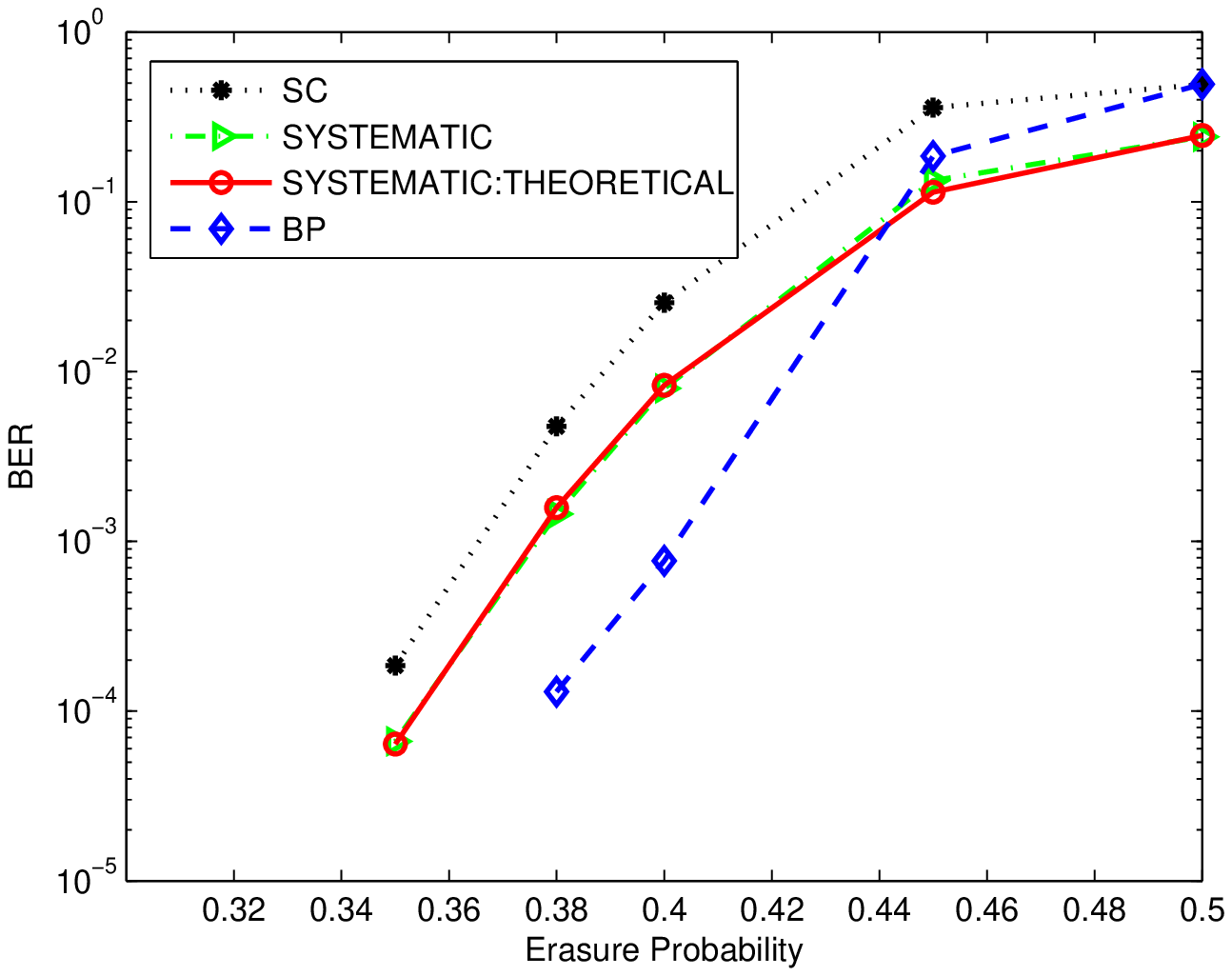}} \par}
\caption{BER for $n=12$, $R=1/2$ in BEC channel. }
\label{fig_ber_bec_n_12}
\end{figure*}

We observe the same phenomenon in Fig.~\ref{fig_ber_bec_n_12} as in Fig.~\ref{fig_ber_bec_n_10} for $N=10^{12}$ and $R=1/2$.
The curves in Fig.~\ref{fig_ber_bec_n_12} have the same style and labels as Fig.~\ref{fig_ber_bec_n_10}.
The coupling coefficient is set the same as the case $N=10$ and $R=1/2$. Again, the
simulated systematic gain embedded in the BER of systematic polar codes matches that calculated using the composite
set $\mathcal{S}$.

\begin{figure*}
{\par\centering
\resizebox*{3.0in}{!}{\includegraphics{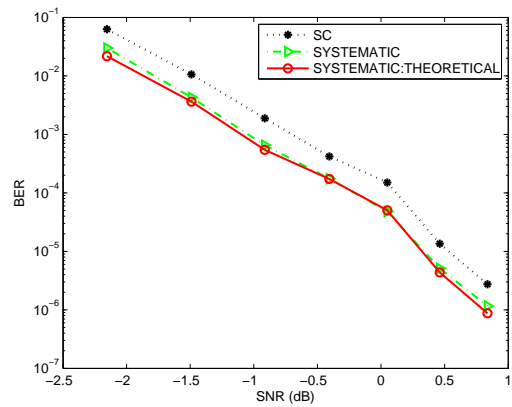}} \par}
\caption{BER for $n=10$, $R=1/4$ in AWGN channel. }
\label{fig_ber_awgn_n_10}
\end{figure*}

Showing in Fig.~\ref{fig_ber_awgn_n_10} is the BER for $N=10$ and $R=1/4$ in the AWGN channel.
The composite set is $\mathcal{S} = \mathcal{A}$. The coupling
coefficient is set in the following way:
for SNR smaller than -1.5 dB, $\beta = 0.3$; for SNR larger than -1.5 dB, $\beta = 0.5$. The systematic
gain calculated from the model in (\ref{eq_st1})(\ref{eq_st2}) matches that with the simulations, showing that the composite
model in (\ref{eq_st1})(\ref{eq_st2}) can also be used for AWGN channels.

\begin{figure*}
{\par\centering
\resizebox*{3.0in}{!}{\includegraphics{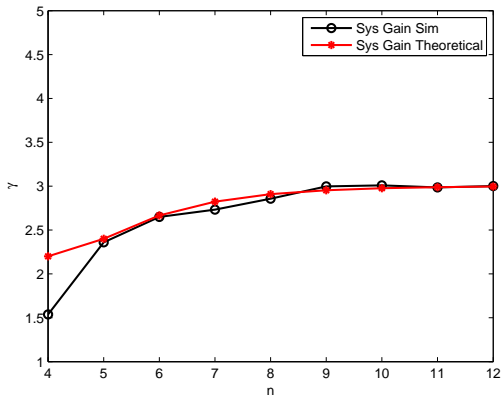}} \par}
\caption{Systematic gain $\gamma$ for $R=1/2$ in BEC channels with different block lengths. The erasure probability is set
the same for all block lengths as 0.4. }
\label{fig_sys_gain_n_4_12}
\end{figure*}

From Fig.~\ref{fig_ber_bec_n_10} to Fig.~\ref{fig_ber_awgn_n_10}, we see that systematic polar codes perform consistently
better than non-systematic polar codes, echoing the results in \cite{arikan_icl11}. The systematic gain  for
different block lengths is shown in Fig.~\ref{fig_sys_gain_n_4_12}.
The underlying channel $W$ is a BEC channel with an
erasure probability 0.4.
The gain represented by the circled line (with a legend 'Sys Gain Sim') is simulated.
The gain shown by the stared line (with a legend 'Sys Gain Theoretical') is calculated using (\ref{eq_mean_gamma}).
The systematic gain calculated using (\ref{eq_mean_gamma}) is accurate when $N$ is large.
It's seen from Fig.~\ref{fig_sys_gain_n_4_12} that the systematic
gain increases with the increase of the block lengths but saturates at around $\gamma = 3$ when $N \ge 2^9$.
This coincides with the simulation
results in Fig.~\ref{fig_ber_bec_n_10} \mytilde~\ref{fig_ber_awgn_n_10}.
The saturating nature of the systematic gain can be seen
from the composite set $\mathcal{S}$: With a fixed code rate $R$, as the block length $N$ increases, the cardinality of
$\mathcal{S}$ also increases. So the increase in the error-correction capability of the systematic polar codes is counteracted
by the increase in the number of error bits, rendering the systematic gain to reach a limit.

\section{Conclusion}\label{sec_con}
In this paper, we analyze the error performance of systematic polar codes with the SC-EN decoding.
Through the analysis of the generating matrix of polar codes, the encoding process of
systematic polar codes is simplified. We use a parameter, the systematic gain, to
characterize the performance of systematic polar codes as compared with the non-systematic
polar codes. From the study of the basic error patterns and the first
error distribution of the SC decoding, the information bits are divided into two regions and the probability
of errors in each region is provided. To further use the properties of these two regions, we propose
a composite model to approximate the mean effect
of the error events in the SC decoding. Using this composite model, the systematic gain can be calculated.
Numerical results are provided and our models are verified in the paper.
Systematic polar codes are shown to be around 3 times  better than non-systematic polar codes in terms of the BER performance
with large block lengths.

\appendix[Proof of the Error Patterns]
We provide the proof of the error pattern in Fig.~\ref{fig_error_pattern}-e. Let's assume the two bits at the input to the Z-section is $u_1$ and $u_2$. The output is then $x_1=u_1 \oplus u_2$ and $x_2=u_2$. In this pattern, the LR value of $x_1$ is
incorrect, namely $LR(x_1) = LR(u_1 \oplus u_2 \oplus 1)$. In estimating $u_1$, we have
\begin{equation}
LR(\hat{u}_1) = \frac{1+LR(u_1 \oplus u_2 \oplus 1)*LR(u_2)}{LR(u_1 \oplus u_2 \oplus 1)+LR(u_2)}
\end{equation}
Compared with the true estimation
\begin{equation}
LR({u}_1) = \frac{1+LR(u_1 \oplus u_2)*LR(u_2)}{LR(u_1 \oplus u_2)+LR(u_2)}
\end{equation}
it's readily seen that $\hat{u}_1 = u_1 \oplus 1$. Therefore the LR value of the variable node $u_1$  is incorrect,
as indicated by a X in the upper left node in Fig.~\ref{fig_error_pattern}-e.

After obtaining the estimation of $\hat{u}_1$, the LR value of bit $u_2$ is given by
\begin{equation}\label{eq_lru2}
LR(\hat{u}_2) = LR(u_2)* LR(u_1\oplus u_2 \oplus 1)^{1-2\hat{u}_1} \\
\end{equation}
Substituting $\hat{u}_1 = u_1 \oplus 1$ into (\ref{eq_lru2}) and using the fact that $LR(u_1 \oplus 1) = LR(u_1)^{-1}$, we
obtain the following
\begin{equation}\label{eq_lru2_2}
LR(\hat{u}_2) = LR(u_2) * LR(u_1\oplus u_2)^{-1 + 2(u_1 \oplus 1)}
\end{equation}
Again, comparing with the true estimation of $u_2$
\begin{equation}\label{eq_lru2_3}
LR({u}_2) = LR(u_2) * LR(u_1\oplus u_2)^{1 - 2u_1}
\end{equation}
we can verify that (\ref{eq_lru2_2}) and (\ref{eq_lru2_3}) are equivalent, meaning the estimation of $\hat{u}_2$ is the true
estimation, which is the lower left node in  Fig.~\ref{fig_error_pattern}-e.

\bibliographystyle{IEEEtran}
\bibliography{../ref_polar}

\begin{thebibliography}{10}
\providecommand{\url}[1]{#1}
\csname url@samestyle\endcsname
\providecommand{\newblock}{\relax}
\providecommand{\bibinfo}[2]{#2}
\providecommand{\BIBentrySTDinterwordspacing}{\spaceskip=0pt\relax}
\providecommand{\BIBentryALTinterwordstretchfactor}{4}
\providecommand{\BIBentryALTinterwordspacing}{\spaceskip=\fontdimen2\font plus
\BIBentryALTinterwordstretchfactor\fontdimen3\font minus
  \fontdimen4\font\relax}
\providecommand{\BIBforeignlanguage}[2]{{%
\expandafter\ifx\csname l@#1\endcsname\relax
\typeout{** WARNING: IEEEtran.bst: No hyphenation pattern has been}%
\typeout{** loaded for the language `#1'. Using the pattern for}%
\typeout{** the default language instead.}%
\else
\language=\csname l@#1\endcsname
\fi
#2}}
\providecommand{\BIBdecl}{\relax}
\BIBdecl

\bibitem{arikan_iti09}
E.~Arikan, ``{Channel Polarization: A Method for Constructing
  Capacity-Achieving Codes for Symmetric Binary-Input Memoryless Channels},''
  \emph{IEEE Transactions on Information Theory}, vol.~55, no.~7, pp.
  3051--3073, 2009.

\bibitem{sasoglu_09}
E.~Sasoglu, E.~Telatar, and E.~Arikan, ``{Polarization for Arbitrary Discrete
  Memoryless Channels},'' Online: \url{http://arxiv.org/pdf/0908.0302v1.pdf}.

\bibitem{mori_itw10}
R.~Mori and T.~Tanaka, ``{Non-Binary Polar Codes using Reed-Solomon Codes and
  Algebraic Geometry Codes},'' in \emph{IEEE Information Theory Workshop
  (ITW)}, 2010, pp. 1--5.

\bibitem{pradhan_allerton11}
A.~G. Sahebi and S.~S. Pradhan, ``{Multilevel Polarization of Polar Codes Over
  Arbitrary Discrete Memoryless Channels},'' in \emph{2011 49th Annual Allerton
  Conference on Communication, Control, and Computing (Allerton)}, September
  2011, pp. 1718--1725.

\bibitem{mori_isit09}
R.~Mori and T.~Tanaka, ``{Performance and Construction of Polar codes on
  Symmetric Binary-Input Memoryless Channels},'' in \emph{IEEE International
  Symposium on Information Theory}, June 2009, pp. 1496--1500.

\bibitem{telatar_isit11}
R.~Pedarsani, S.~Hassani, I.~Tal, and I.~Telatar, ``{On the Construction of
  Polar Codes},'' in \emph{IEEE International Symposium on Information Theory
  Proceedings (ISIT)}, 2011, pp. 11--15.

\bibitem{trifonov_itc12}
P.~Trifonov, ``{Efficient Design and Decoding of Polar Codes},'' \emph{IEEE
  Transactions on Communications}, vol.~60, no.~11, pp. 3221--3227, November
  2012.

\bibitem{vardy_polar}
I.~Tal and A.~Vardy, ``{How to Construct Polar Codes},'' Online:
  \url{http://arxiv.org/abs/1304.3850}.

\bibitem{korada_iti10}
S.~B. Korada, E.~Sasoglu, and R.~Urbanke, ``{Polar Codes: Characterization of
  Exponent, Bounds, and Constructions},'' \emph{IEEE Transactions on
  Information Theory}, vol.~56, no.~12, pp. 6253--6264, December 2010.

\bibitem{telatar_itw10}
E.~Abbe and I.~Telatar, ``{MAC Polar Codes and Matroids},'' in
  \emph{Information Theory and Applications Workshop (ITA)}, Jan 2010, pp.
  1--8.

\bibitem{vardy_iti11}
H.~Mahdavifar and A.~Vardy, ``{Achieving the Secrecy Capacity of Wiretap
  Channels Using Polar Codes},'' \emph{IEEE Transactions on Information
  Theory}, vol.~57, no.~10, pp. 6428--6443, October 2011.

\bibitem{telatar_iti12}
E.~Abbe and E.~Telatar, ``{Polar Codes for the m-User Multiple Access
  Channel},'' \emph{IEEE Transactions on Information Theory}, vol.~58, no.~8,
  p. 5437Ð5448, 2012.

\bibitem{arikan_isit09}
E.~Arikan and I.~Telatar, ``{On the Rate of Channel Polarization},'' in
  \emph{IEEE International Symposium on Information Theory (ISIT)}, 2009, pp.
  1493--1495.

\bibitem{urbanke_isit10}
S.~H. Hassani and R.~Urbanke, ``{On the Scaling of Polar Codes: I. The Behavior
  of Polarized Channels},'' in \emph{IEEE International Symposium on
  Information Theory Proceedings (ISIT)}, June 2010, pp. 874--878.

\bibitem{mori_isit10}
T.~Tanaka and R.~Mori, ``{Refined Rate of Channel Polarization},'' in
  \emph{IEEE International Symposium on Information Theory}, June 2010, pp.
  889--893.

\bibitem{hassani_iti13}
S.~Hassani, R.~Mori, T.~Tanaka, and R.~Urbanke, ``{Rate-Dependent Analysis of
  the Asymptotic Behavior of Channel Polarization},'' \emph{Information Theory,
  IEEE Transactions on}, vol.~59, no.~4, pp. 2267--2276, 2013.

\bibitem{urbanke_isit09}
N.~Hussami, S.~Korada, and R.~Urbanke, ``{Performance of Polar Codes for
  Channel and Source Coding},'' in \emph{IEEE International Symposium on
  Information Theory (ISIT)}, June 2009, pp. 1488--1492.

\bibitem{eslami_allerton10}
A.~Eslami and H.~Pishro-Nik, ``{On Bit Error Rate Performance of Polar Codes in
  Finite Regime},'' in \emph{48th Annual Allerton Conference on Communication,
  Control, and Computing (Allerton)}, 2010, pp. 188--194.

\bibitem{arikan_icl08}
E.~Arikan, ``{A Performance Comparison of Polar Codes and Reed-Muller codes},''
  \emph{IEEE Communications Letters}, vol.~12, no.~6, pp. 447--449, 2008.

\bibitem{guo_isit14}
J.~Guo, M.~Qin, A.~G. i~Fabregas, and P.~H. Siegel, ``{Enhanced Belief
  Propagation Decoding of Polar Codes through Concatenation},'' in \emph{2014
  IEEE International Symposium on Information Theory Proceedings (ISIT)}, 2014,
  pp. 2987 -- 2991.

\bibitem{barry_icc13}
U.~U. Fayyaz and J.~R. Barry, ``{Polar Codes for Partial Response Channels},''
  in \emph{2013 IEEE International Conference on Communications (ICC)}, 2013,
  pp. 4337 -- 4341.

\bibitem{vardy_isit11}
I.~Tal and A.~Vardy, ``{List Decoding of Polar Codes},'' in \emph{2011 IEEE
  International Symposium on Information Theory Proceedings (ISIT)}, July 2011,
  pp. 1--5.

\bibitem{niu_itc13}
K.~Chen, K.~Niu, and J.~Lin, ``{Improved Successive Cancellation Decoding of
  Polar Codes},'' \emph{IEEE Transactions on Communications}, vol.~61, no.~8,
  pp. 3100--3107, August 2013.

\bibitem{eslami_isit11}
A.~Eslami and H.~Pishro-Nik, ``{A Practical Approach to Polar Codes},'' in
  \emph{IEEE International Symposium on Information Theory}, 2011, pp. 16--20.

\bibitem{arikan_icl11}
E.~Arikan, ``{Systematic Polar Coding},'' \emph{IEEE Communications Letters},
  vol.~15, no.~8, pp. 860--862, August 2011.

\end{thebibliography}

\end{document}